\documentclass[
]{ceurart}

\usepackage{listings}
\usepackage{array}
\usepackage{enumitem}
\usepackage{xspace}
\usepackage{tikz}
\usepackage{amsmath}
\usepackage{subcaption}
\usepackage{threeparttable}
\usepackage{pgfplots}

\usetikzlibrary{arrows.meta, positioning, fit, backgrounds, calc}
 \pgfplotsset{compat=1.17}

\definecolor{inputcol}{RGB}{173, 216, 230}
\definecolor{embedcol}{RGB}{144, 238, 144}
\definecolor{blockcol}{RGB}{255, 218, 185}
\definecolor{outputcol}{RGB}{255, 182, 193}
\definecolor{swapitem}{RGB}{235, 210, 250}

\definecolor{coldgray}{HTML}{9a9a9a}
\colorlet{sasrecblue}{blue!70!black}
\colorlet{swaprecorange}{orange!80!black}

\tikzset{
  box/.style={rectangle, rounded corners=4pt, draw=black, line width=0.7pt,
    minimum width=0.8cm, minimum height=0.55cm, align=center, font=\small},
  ibox/.style={box, fill=inputcol},
  sbox/.style={box, fill=swapitem, draw=purple!60},
  ebox/.style={box, fill=embedcol},
  bblock/.style={box, fill=blockcol, minimum width=1.4cm, font=\scriptsize},
  obox/.style={box, fill=outputcol},
  swapblock/.style={rectangle, rounded corners=4pt, draw=purple!55, fill=purple!6,
    line width=0.7pt, minimum width=1.3cm, minimum height=0.55cm,
    align=center, font=\scriptsize\bfseries, text=purple!70},
  condbox/.style={rectangle, rounded corners=4pt, draw=purple!45, fill=purple!4,
    line width=0.7pt, minimum width=3.0cm, minimum height=0.55cm,
    align=center, font=\small, text=purple!75},
  arr/.style={-Stealth, thick, black!65},
  arryes/.style={-Stealth, thick, purple!65},
  arrno/.style={-Stealth, thick, black!65},
  ann/.style={font=\footnotesize\itshape, text=black!50},
  yeslabel/.style={font=\scriptsize\bfseries, text=purple!70, midway, right=1pt},
  nolabel/.style={font=\scriptsize\bfseries, text=black!50, midway, right=1pt},
}
 
\newcommand{\ie}{i.\,e., }

\newif\ifworkinprogress
\workinprogressfalse

\ifworkinprogress
	\newcommand{\ms}[1]{\textcolor{blue}{{[Markus] #1}}}
	\newcommand{\mm}[1]{\textcolor{olive}{{[Marta] #1}}}
	\newcommand{\mr}[1]{\textcolor{green}{{[Matteo] #1}}}
    \newcommand{\ml}[1]{\textcolor{red}{{[Malte] #1}}}
	\newcommand{\da}[1]{\textcolor{orange}{{[Davide] #1}}}
       
\else
    \newcommand{\ms}[1]{}
    \newcommand{\mm}[1]{}
    \newcommand{\mr}[1]{}
    \newcommand{\ml}[1]{}
    \newcommand{\da}[1]{}
       
\fi

\newcommand{\ourapproach}{\text{SwapRec}\xspace} 
\newcommand{\onion}{\text{Onion}\xspace}
\newcommand{\amazon}{\text{Amazon}\xspace}
\newcommand{\movielens}{\text{ML-20M}\xspace}

\newcommand{\multvae}{\text{MultVAE}\xspace}

\newcommand{\als}{\text{ALS}\xspace}
\newcommand{\iknn}{\text{Item-}$k$\text{NN}\xspace}

\newcommand{\bert}{\text{BERT4Rec}\xspace}
\newcommand{\sasrec}{\text{SASRec}\xspace}
\newcommand{\swapbert}{\text{BERT4Rec w. \ourapproach}\xspace}
\newcommand{\swaprec}{\text{SASRec w. \ourapproach}\xspace}

\newcommand{\pswap}{\ensuremath{p_{\mathrm{swap}}}\xspace}
\newcommand{\maxswap}{\ensuremath{M_{\mathrm{swap}}}\xspace}
\newcommand{\hrten}{\text{HR@10}\xspace}
\newcommand{\traininter}{\ensuremath{n_\text{train}}}
\newcommand{\ncold}{\ensuremath{n_\text{cold}}}

\newcolumntype{H}{>{\setbox0=\hbox\bgroup}c<{\egroup}@{}}

\newcommand{\colsep}{0.7cm}

\begin{document}

\copyrightyear{2026}
\copyrightclause{Copyright for this paper by its authors. Use permitted under Creative Commons License Attribution 4.0 International (CC BY 4.0). CEUR Workshop Proceedings (CEUR-WS.org).}

\conference{{DaQuaMRec} @ RecSys 2026: Second International Workshop on Data Quality-Aware Multimodal Recommendation}

\title{SwapRec: 
Warming Up Cold Items Through Training-Time Swaps}


\author[1,2]{Marta Moscati}[%
orcid=0000-0002-5541-4919,
email=marta@usealbatross.ai,
]
\cormark[1]
\address[1]{Albatross AI}
\address[2]{Johannes Kepler University Linz,
  Linz, Austria}

\author[]{Jan Malte Lichtenberg}[%
orcid=0000-0002-7140-2520,
]

\author[1]{Davide Abbattista}[%
orcid=0009-0008-7683-0533,
email=davide@usealbatross.ai,
]

\author[1]{Antonio De Candia}[%
email=antonio@usealbatross.ai,
]

\author[1]{Laura Boggia}[%
orcid=0000-0002-9924-7489,
email=laura@usealbatross.ai,
]

\author[1]{Matteo Ruffini}[%
orcid=0000-0003-0738-2198,
email=matteo@usealbatross.ai,
]

\cortext[1]{Corresponding author.}

\begin{abstract}
  Interactions with cold items negatively impact real-time personalization of ID-based recommender systems. This is because the use of such interactions degrades user preference estimates, whereas excluding cold items from the user profile prevents real-time recommendation updates. In industrial scenarios, one heuristic often applied to address this shortcoming at inference time is to replace, \ie ``swap'', cold-start items by their most similar ``warm'' neighbor, where similarity is inferred from the items' side information.  In this paper, we demonstrate that sequential models, most often used for real-time personalization, are not robust to such swaps, and propose \ourapproach, an approach to address this issue. \ourapproach relies on using the same swap heuristics already at training time. We apply \ourapproach to state-of-the-art models for sequential recommendation and analyze its impact by means of quantitative experiments in three recommendation domains (online shopping, movie, music). The experimental results show that, irrespective of the underlying sequential architecture, our easy-to-implement \ourapproach approach allows for substantially more accurate recommendations when in presence of interactions with cold items, simultaneously leading to a larger percentage of cold items in the recommendation lists. 
\end{abstract}

\begin{keywords}
  Item Cold Start \sep
  Sequential Recommendation \sep
  Real-Time Personalization
\end{keywords}

\maketitle

\section{Introduction}
    Recommender Systems (RS) are nowadays ubiquitous and help users in selecting products from the large catalogs of online providers. Most RS are ID-based and optimize a separate embedding for each catalog item based on behavioral user data. One challenge for ID-based RS is handling cold items, i.e., items that have been interacted by few or no users. Item cold-start negatively impacts both users and content producer. From the user perspective, a click on a cold item can degrade real-time personalization of ID-based RS since a click on an item with no previous interactions typically leads to 
    inaccurate recommendations, 
    but simply disregarding this click prevents real-time updates of recommendations. From the content producer perspective, ID-based RS are unable to surface cold items and, as a consequence, newly added items are never recommended.

\begin{figure}[h]
    \centering
    \includegraphics[width=\linewidth]{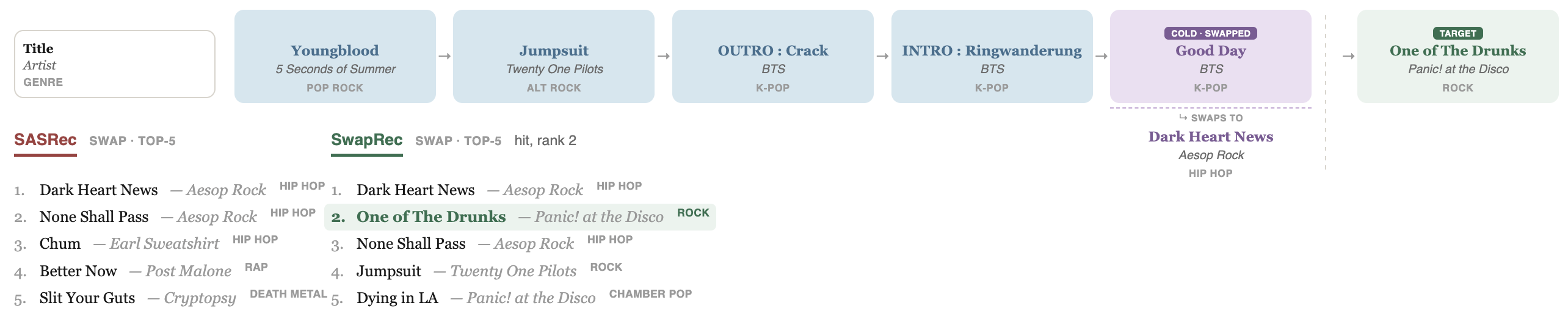}
    \caption{Example of a test sequence from the Music4All-Onion dataset showing that \ourapproach renders recommendation robust to inference-time similarity swaps. The swap map is computed as described in Section~\ref{sec:experiments}.}
    \label{fig:example}
\end{figure}

    These issues of item cold-start are typically addressed with hybrid RS that leverage item content representations alongside behavioral data.  However, these approaches are more complex than models relying solely on behavioral data, require more resources, or a more demanding hyperparameter tuning. 

    As a consequence, industrial RS application choose simpler solutions that replace clicks on a cold item with clicks on the most similar warm item. Figure~\ref{fig:example} shows a real-world example of this approach when applied to music streaming. Blue cards represent interactions with warm items (\textit{Youngblood}, \textit{Jumpsuit}, ...), and the purple card represents an interaction with a cold item (\textit{Good Day}). This item is swapped with its most similar warm catalog item (\textit{Dark Heart News}) shown below it, where similarity is inferred from the audio signal. This swapped sequence can now be seamlessly used as input to ID-based RS, since it only constitutes of warm items.
    
    Approaches based on this heuristic are widely employed since they 
    allow using the already-deployed RS without the need to switch to a content-based approach or to train an additional component for semantic ID generation. However, swapping items at inference time relies on the implicit assumption that the underlying RS is robust to such swaps. This assumption is unrealistic, since semantically similar warm and cold items are likely to be interacted with by different users or within different sequences, for instance because the items were added to the catalog at different times, or because a less popular item occurs in fewer interaction sequences. This leads to a decrease in accuracy of ID-based RS when users interact with cold items and those are swapped with similar ones. In the example shown in Figure~\ref{fig:example}, this effect is reflected in the fact that the top-5 recommendation list of the ID-based RS SASRec does not include the next item the user will interact with (Target item, green card, \textit{One of The Drunks}).   

    In this paper we propose to \textbf{make ID-based  RS robust to inference-time item swaps by introducing such swaps already during training} and develop \ourapproach, a new training strategy that makes ID-based RS robust to interactions with cold items. In the example shown in Figure~\ref{fig:example}, we see that the next item interacted with by the user is included in the top-5 recommendation list of \ourapproach. 
    
    In order to quantitatively evaluate our proposed \ourapproach approach, we pose the following research questions (RQ):
    \begin{description}
        \item[RQ1: Overall Model Performance.] How does \ourapproach compare to standard training strategies of ID-based RS in terms of overall recommendation accuracy?
        \item[RQ2: Robustness to Cold-Item Interactions.] Does \ourapproach 
        mitigate the performance deterioration of sequential ID-based RS when interactions with cold items are present in the input sequence?
        \item[RQ3: Cold Items Recommended.] Do  \ourapproach's recommendations  surface cold items more often in recommendation lists, compared to ID-based RS?
    \end{description}
    
    Compared to previous works addressing item cold-start, \ourapproach can be seamlessly integrated into ID-based RS, without the need to change the model architecture nor to train additional external components, hence allowing to address item cold-start without additional model complexity. 
    
    We test \ourapproach on top of well-established transformer-based and ID-based sequential RS. 
    Our experiments in three recommendation domains (movie, music, and online shopping) show that, irrespective of the underlying sequential architecture, our easy-to-implement \ourapproach approach allows for substantially more accurate recommendations in presence of interactions with cold items and surfaces more cold items; these results, combined with \ourapproach's simplicity, make \ourapproach a favorable solution to item cold-start for users, content producers, as well as platform providers. 

\section{Related Work}
    Item cold-start, \ie providing accurate recommendations even for items that have not been interacted with by any user, is a major challenge for RS~\cite{ricci2022rshandbook}. Several works~\cite{ganhoer_moscati2024sibrar,wei2021clcrec,volkovs2017dropoutnet,malitesta2024formal,Ganhoer2025TORS_SiBraR} showed that this challenging situation can be tackled with multimodal RS, that complement behavioral data such as user clicks, with multiple item representations. %
    
    In ID-based RS, the negative impact of clicks on cold item originates from the fact that IDs of items with no clicks in the training data 
    do not get updated during training. To address these limitations, previous works proposed methods to infer items behavioral IDs from items side information ~\cite{lichtenberg2025daquamrec,lichtenberg2025denserec,pembec2025letitgo}  
    or to replace behavioral IDs with (pre-constructed) IDs that are ``semantic'', \ie representative of the item side information~\cite{singh2024semantic_ids,rajput2023semantic_ids,yang2025liger}.

    Therefore, in real-world applications, where ID-based sequential RS are most often used for real-time personalization~\cite{koneru2024sasrec_zdf,koneru2025sasrec_zdf,klenitskiy2023sasrec,lichtenberg2025denserec}, companies often choose to handle sequences including cold items with simple heuristics. These simpler approaches are the focus of our work. 
    
    To give concrete examples, providers in the domain of housing marketplace~\cite{grbovic2018real_time_personalization_airbnb} 
    and e-commerce~\cite{seol2024proxy_alibaba} at inference time use as proxies for cold-start item embeddings the average embeddings of the most similar ``warm'' items, defining similarity in terms of side information. In other terms, they ``swap'' cold items with their most similar warm ones. More recent approaches leverage similar heuristics and substitute the item encoding layer of sequential RS with the weighted average of proxies based on item side information~\cite{seol2024proxy_alibaba}.

    Therefore, a model trained solely on behavioral data struggles in learning that semantically similar items should be treated similarly~\cite{collins2026ensrec}, leading to a performance deterioration when the swap heuristic is applied to cold items at inference time. 
    
    Compared to previous works leveraging multimodal item representations to address item cold-start scenarios, \ourapproach can be seamlessly integrated into ID-based RS, without the need to change the model architecture nor to train additional external components. Compared to previous works which apply inference-time swaps to replace cold-start item embeddings, \ourapproach leverages the content-based similarity swaps already during training; this renders ID-based recommendations more robust to such swaps for item cold-start and allows surfacing a larger portion of catalog items.

\section{Methodology}

\label{sec:methodology}
    \subsection{Preliminaries} Given the set of users $\mathcal{U}$ and of items 
    $\mathcal{I}$
    , for each $u\in \mathcal{U}$, sequential RS represent the user $u$ interaction history as chronologically ordered sequence of $\ell$ interactions $\mathbf{s}^u = (s_1, \ldots, s_\ell
    ), s_k\in \mathcal{I}$
    . 
    
    Each item $i \in \mathcal{I}$, will occur 
    $\traininter$ times in the training data; this value therefore distinguishes cold items from warm ones. Items for which $\traininter = 0$ correspond to a \textit{strict} cold start; items for which $\traininter < \ncold$ are considered cold.\footnote{We refer the reader to Sec.~\ref{sec:experiments} for the choice of $\ncold$.} 
    
    Each $i \in \mathcal{I}$ is associated with side information (e.g.\ description, image) encoded in a space with a notion of similarity $\mathrm{sim}(i, j)$  between two items $i,j \in \mathcal{I}$. 
    This metric space induces a nearest-neighbor (NN) map \begin{equation}\phi(i) = \operatorname*{arg\,max}_{j \in \mathcal{I} \setminus \{i\}} \mathrm{sim}(i, j)  
    \end{equation} 
    associating each item $i$ with its NN. 
    \subsection{\ourapproach}\ourapproach is based on the intuition that ID-based RS trained solely on behavioral data struggle to encode semantic similarities~\cite{collins2026ensrec}. To solve this, \ourapproach 
    treats semantically similar items in the same way during training. 
    
    The goal is to make item embeddings robust to swaps between semantically similar items. To do so, the information encoded in the embeddings at the end of \ourapproach training combines behavioral patterns and item semantics, injecting item side information into behavioral ID embeddings.
    
    \paragraph{\ourapproach{} Training. }During training, \ourapproach{} leverages the NN map $\phi$ to swap items in the sequence with the most similar ones.  
    
    For each position $k$ in $\mathbf{s}^u$, we independently randomly sample $r_k \sim \mathcal{U}(0,1)$ and substitute  $s_k$ with $\tilde{s}_k$, where
    \[
      \tilde{s}_k =
      \begin{cases}
        \phi(s_k) & \text{if } r_k < \pswap
                    \;\text{ and }\;
                    \displaystyle\sum_{i=1}^{k} \mathbf{1}[\tilde{s}_i \neq s_i] < M_{\mathrm{swap}}, \\
        s_k       & \text{otherwise.}
      \end{cases}
    \]
    
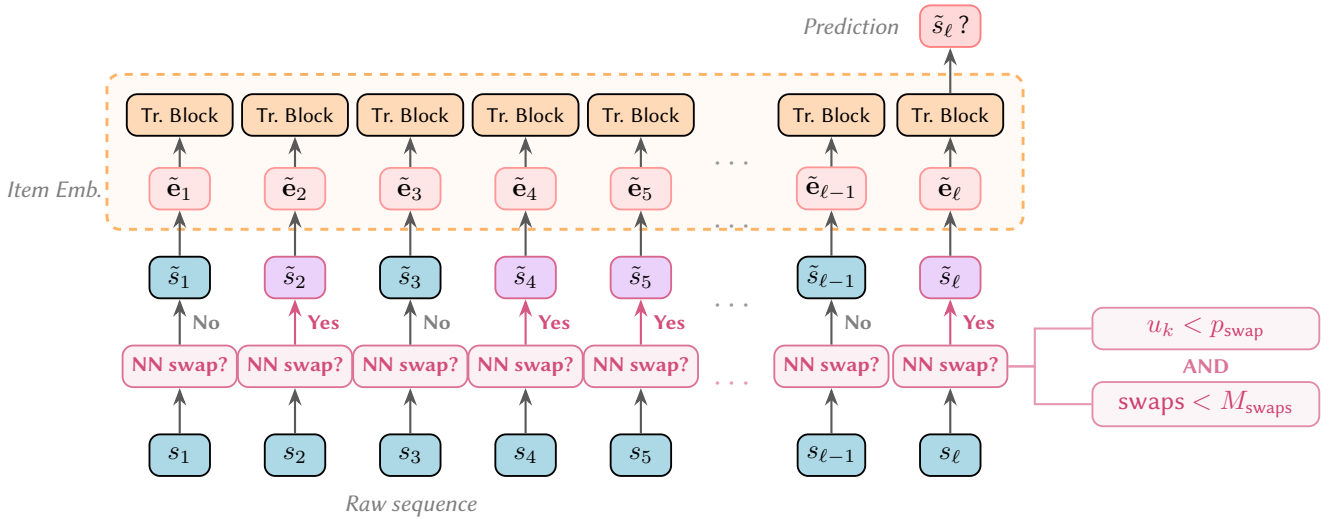
\begin{figure*}[ht]
\centering
\begin{tikzpicture}[node distance=0.5cm]
 
  \node[ibox] (r1)  {$s_1$};
  \node[ibox, right=\colsep of r1] (r2)  {$s_2$};
  \node[ibox, right=\colsep of r2] (r3)  {$s_3$};
  \node[ibox, right=\colsep of r3] (r4)  {$s_4$};
  \node[ibox, right=\colsep of r4] (r5)  {$s_5$};
  \node[right=\colsep of r5, font=\large, text=black!40] (rd){};
  \node[ibox, right=\colsep of rd]  (rn)  {$s_{\ell - 1}$};
  \node[ibox, right=\colsep of rn]  (rn1) {$s_{\ell}$};
  \node[ann, below=0.12cm of r3] {Raw sequence};
 
  \node[swapblock, above=0.6cm of r1]  (m1)  {NN swap?};
  \node[swapblock, above=0.6cm of r2]  (m2)  {NN swap?};
  \node[swapblock, above=0.6cm of r3]  (m3)  {NN swap?};
  \node[swapblock, above=0.6cm of r4]  (m4)  {NN swap?};
  \node[swapblock, above=0.6cm of r5]  (m5)  {NN swap?};
  \node[above=0.6cm of rd, font=\large, text=purple!40] (md) {$\cdots$};
  \node[swapblock, above=0.6cm of rn]  (mn)  {NN swap?};
  \node[swapblock, above=0.6cm of rn1] (mn1) {NN swap?};
 
  \foreach \a/\b in {r1/m1,r2/m2,r3/m3,r4/m4,r5/m5,rn/mn,rn1/mn1}
    \draw[arr] (\a.north) -- (\b.south);
 
  \node[condbox, right=1.1cm of mn1, yshift= 0.5cm] (cond1) {$u_k < p_{\text{swap}}$};
  \node[condbox, right=1.1cm of mn1, yshift=-0.5cm] (cond2) {swaps $< M_{\text{swaps}}$};
  \coordinate (cmid) at ($(mn1.east)+(0.35cm,0)$);
  \draw[purple!40, thick] (mn1.east) -- (cmid) |- (cond1.west);
  \draw[purple!40, thick] (mn1.east) -- (cmid) |- (cond2.west);
  \node[font=\scriptsize\bfseries, text=purple!55,
        at={($(cond1.south)!0.5!(cond2.north)$)}] {AND};
 
  \node[ibox, above=0.6cm of m1]  (t1)  {$\tilde{s}_1$};
  \node[sbox, above=0.6cm of m2]  (t2)  {$\tilde{s}_2$};
  \node[ibox, above=0.6cm of m3]  (t3)  {$\tilde{s}_3$};
  \node[sbox, above=0.6cm of m4]  (t4)  {$\tilde{s}_4$};
  \node[sbox, above=0.6cm of m5]  (t5)  {$\tilde{s}_5$};
  \node[above=0.6cm of md, font=\large, text=black!40] (td) {$\cdots$};
  \node[ibox, above=0.6cm of mn]  (tn)  {$\tilde{s}_{\ell - 1}$};
  \node[sbox, above=0.6cm of mn1] (tn1) {$\tilde{s}_{\ell}$};
 
  \draw[arrno]  (m1.north)  -- (t1.south)  node[nolabel]  {No};
  \draw[arryes] (m2.north)  -- (t2.south)  node[yeslabel] {Yes};
  \draw[arrno]  (m3.north)  -- (t3.south)  node[nolabel]  {No};
  \draw[arryes] (m4.north)  -- (t4.south)  node[yeslabel] {Yes};
  \draw[arryes] (m5.north)  -- (t5.south)  node[yeslabel] {Yes};
  \draw[arrno]  (mn.north)  -- (tn.south)  node[nolabel]  {No};
  \draw[arryes] (mn1.north) -- (tn1.south) node[yeslabel] {Yes};
 
  \node[ebox, above=0.6cm of t1, fill=red!10, draw=red!40] (e1) {$\tilde{\mathbf{e}}_1$};
  \node[ebox, above=0.6cm of t2, fill=red!10, draw=red!40] (e2) {$\tilde{\mathbf{e}}_2$};
  \node[ebox, above=0.6cm of t3, fill=red!10, draw=red!40] (e3) {$\tilde{\mathbf{e}}_3$};
  \node[ebox, above=0.6cm of t4, fill=red!10, draw=red!40] (e4) {$\tilde{\mathbf{e}}_4$};
  \node[ebox, above=0.6cm of t5, fill=red!10, draw=red!40] (e5) {$\tilde{\mathbf{e}}_5$};
  \node[above=0.6cm of td, font=\large, text=black!40] (ed) {$\cdots$};
  \node[ebox, above=0.6cm of tn,  fill=red!10, draw=red!40] (en)  {$\tilde{\mathbf{e}}_{\ell - 1}$};
  \node[ebox, above=0.6cm of tn1, fill=red!10, draw=red!40] (en1) {$\tilde{\mathbf{e}}_{\ell}$};
  \node[ann, left=0.5cm of e1] {Item Emb.};
 
  \foreach \a/\b in {t1/e1,t2/e2,t3/e3,t4/e4,t5/e5,tn/en,tn1/en1}
    \draw[arr] (\a.north) -- (\b.south);
 
  \node[bblock, above=0.4cm of e1] (b1) {Tr. Block};
  \node[bblock] (b2)  at (e2  |- b1) {Tr. Block};
  \node[bblock] (b3)  at (e3  |- b1) {Tr. Block};
  \node[bblock] (b4)  at (e4  |- b1) {Tr. Block};
  \node[bblock] (b5)  at (e5  |- b1) {Tr. Block};
  \node[above=0.4cm of ed, font=\large, text=black!40] (bd) {$\cdots$};
  \node[bblock]  (bn)  at (en  |- b1) {Tr. Block};
  \node[bblock]  (bn1) at (en1 |- b1) {Tr. Block};
 
  \foreach \a/\b in {e1/b1,e2/b2,e3/b3,e4/b4,e5/b5,en/bn,en1/bn1}
    \draw[arr] (\a.north) -- (\b.south);
 
  \begin{scope}[on background layer]
    \node[draw=orange!60, dashed, line width=1.1pt, fill=orange!4,
          rounded corners=5pt, fit=(b1)(bn1)(e1)(en1)(bd), inner sep=0.25cm,
          label={},
          ]
      (tblock) {};
  \end{scope}
 
  \node[obox, above=0.6cm of bn1, fill=red!15, draw=red!50] (pn1) {$\tilde{s}_{\ell}$\,?};
  \node[ann, left=0.15cm of pn1] {Prediction};
 
  \draw[arr] (bn1.north) -- (pn1.south);
 
\end{tikzpicture}%
\caption{\ourapproach applied to \sasrec. Input sequence items $s_k$ are either left unchanged or swapped (both as target and input) with $\phi(s_k)$,
 with probability $p_{\text{swap}}$ and
if the number of total swaps has not exceeded $M_{\text{swaps}}$.}
\label{fig:swaprec}
\end{figure*}

    The swap-augmented sequence $\tilde{\mathbf{s}} = (\tilde{s}_1, \ldots, \tilde{s}_\ell
    )$ is used to train the backbone sequential RS. Notice that the item swaps affect both the training targets and the input sequences. Figure~\ref{fig:swaprec} provides an overview of \ourapproach when applied to the transformer-based architecture \sasrec~\cite{kang2018sasrec}. 
    
    To provide an intuition behind \ourapproach{}'s training, the two parameters $\pswap$ and $\maxswap$ adjusts the amount of training swaps. Each sequence item is swapped with probability $\pswap$ and only if less than $\maxswap$ items were previously swapped for the same sequence. 

    Notice that these swaps occur irrespective of the number of training interactions   $\traininter$ of the target item in the sequence. Since cold items have fewer interactions in the training data, they are less likely to occur in 
    unswapped training sequences. Therefore 
    swaps are likely to exchange warm items with cold ones. This implies more ID embedding updates for cold items, when those are selected as items to be swapped with.

     Enforcing $M_\text{swap}$ as maximum number of swaps allows to control the extent to which \ourapproach modifies $u$'s true behavioral data. Since all subsequences of $\mathbf{\tilde{s}}^u$ are used for training (\ie each $s_k, k> 1$ is used once as target), $M_\text{swap}$ also implies that swaps occur earlier in the training sequences, further increasing the number of ID embedding updates for cold items. 

    Both \pswap and \maxswap are treated as hyperparameters, as discussed in the  Section on the Experimental Setup.    
    \paragraph{\ourapproach Inference. }At inference time, \ourapproach swaps cold item $s_k$ in the input sequence with its most similar one  according to the NN map $\phi(s_k)$. Notice that at inference the swap does \textit{not} affect the target item, \ie the one to be recommended.
    
\section{Experimental Setup}
    \label{sec:experiments}
    \subsection{Datasets and Preprocessing}
    We carry out experiments on three recommendation domains: music streaming, online shopping, and movie ratings. %
        
        \textit{Music4All-Onion}~\cite{moscati2021onion} is a multimodal dataset for music recommendation based on the tracks of Music4All~\cite{santana2020music4all} and on the interaction data of LFM-2b~\cite{lfm2b}. Following Li et al.~\cite{li2026diffusion}, we restrict the dataset to the interactions happening in 2018. As item side information, we use the representations of the audio signals extracted with a pre-computed instance of MusiCNN~\cite{pons2018atscale,pons2019musicnn} and provided by Peintner et al.~\cite{peintner2025srgnnemo}. 
        
        \textit{\amazon}~\cite{ni2019amazonreviews,hou2024amazonreviews} is a dataset of user reviews for purchases on the Amazon online shopping platform. We use the All\_Beauty reviews of the 2023 version of the dataset, with the concatenated text of product title and description as side information. Following Spillo et al.~\cite{spillo2023combining}, the concatenated text is embedded using the pre-trained all-MiniLM-L6-v2 SentenceTransformer model~\cite{reimers2019sentence_bert} from HuggingFace.\footnote{\url{https://huggingface.co/sentence-transformers/all-MiniLM-L6-v2}}

        \textit{\movielens} is the 20-millions version of the MovieLens movie ratings dataset~\cite{harper2015movielens20m}. This dataset was recently enriched with multimodal item side information by Spillo et al.~\cite{spillo2026movielens}. 
        As side information, we take the movie IMDB\footnote{\url{https://www.imdb.com/}} plot, encoded with a pre-trained all-mpnet-base-v210 SentenceTransformer model~\cite{reimers2019sentence_bert}, as provided by Spillo et al.~\cite{spillo2026movielens}. 

        \paragraph{Preprocessing. }
        Following best practices in sequential recommendation~\cite{gusak2025time_to_split,gusak2025recsys_dish_served_warm}, we split the data using a global temporal split strategy, requiring $20\%$ of the interactions in the test set. For test interactions, the last item of each sequence is treated as target, while the preceding interactions of the same sequence (including those at timestamps preceding the global split timestamp), form the input. This setup better reflects real-world usage and prevents leakage of future data~\cite{gusak2025time_to_split}. 
        
        We set the maximum sequence length to $\ell=40$ and partition longer sequences into chunks of length $40$, using all chunks occurring at timestamps preceding the global split timestamp for training. This allows us to fully exploit the available training data without discarding interactions. 
        
        We apply iterative $k$-core filtering to train interactions; since we are interested in item cold-start scenarios, we reduce the typical~\cite{Melchiorre2021RSFairness} value of $k=5$ to $k=2$. We use cosine similarity to compute the NN map $\phi$.  
    \begin{center}
    \begin{table}[!htb]
        \centering
        \begin{tabular}{l|rrrrHHHH}
                & \# users & \# items & \# interactions  & Item Side Information&sparsity &Text&Audio &Image \\ 
            \midrule
            \onion
            &   20{,}455  &   53{,}377  &17{,}285{,}525 &Track audio (MusiCNN~\cite{pons2018atscale,pons2019musicnn})  & &768&50; 100; 4{,}800&4{,}096\\
            \amazon
            & 12,594	&7,105&	500,095& Product title and description  (all-MiniLM-L6-v2~\cite{reimers2019sentence_bert})&&&&\\
            \movielens
            &	112,329	& 11,687	&19,919,743& Movie plot (all-mpnet-base-v210~\cite{reimers2019sentence_bert,spillo2026movielens})
 &&&&\\
        \end{tabular}
        
        \caption{Summary of datasets after pre-processing.}
        \label{tab:dataset-overview}
    \end{table}
    \end{center}




    \subsection{Evaluation}
    \paragraph{RQ1: Overall Model Performance.} We evaluate the overall accuracy of RS on Hit-Rate for lists of $10$ recommendations (\hrten), as commonly done for sequential recommendation~\cite{klenitskiy2024does_it_look_sequential,gusak2025time_to_split,gusak2025recsys_dish_served_warm}, reporting mean and standard error (SE) over test sequences. \\
    \paragraph{RQ2: Robustness to Cold-Item Clicks.} RQ2 focuses on the negative impact of clicks on cold items, \ie from users' perspective. To test model robustness to item cold-start scenarios, as well as to heuristics used to address those situations, we split evaluation sequences according to the number \traininter of train interactions of the last item before the target one. We focus exclusively on the last item since for sequential models the latest interaction is expected to influence recommendations more substantially than earlier interactions~\cite{kang2018sasrec,oh2024recency}. 
    For both the full set of test sequences and test sequences ending with cold items ($\traininter < \ncold$), we evaluate model performance with and without swapping the last item in the input sequence. 

    Additionally, to mimic the strict-cold-start scenarios, we compare models performance when treating the last input item in the following ways.
    \begin{description}
        \item[No swap:] the item is kept as-is. 
        \item[Swap:] the item is swapped with its NN according to $\phi$.
        \item[Init:] the item is assigned a randomly-initialized token embedding, with the same initialization of the underlying ID-based RS, hence mimicking strict cold-start. 
        \item[Drop:] the item is dropped entirely from the input sequence, another technique often used to address strict-item-cold-start scenarios in sequential recommendation.
    \end{description}
    Notice that the evaluations in the last settings (drop), requires input test sequences to consist of at least two items.  For input test sequences consisting of a single item,  dropping it would leave the model without interactions to compute recommendations on, i.e., in a user cold-start scenario, which is beyond the scope of the current work. We therefore limit the ``drop'' model comparison for cold-start to input sequences of at least two items.

    \paragraph{RQ3: Cold Items Recommended.} RQ3 focuses on the negative impact of item cold-start on content producers, \ie the underexposure of cold items in recommendation lists. To compare the extent to which models are able to represent cold items in recommendations, we measure the empirical cumulative distribution function (CDF) of top-$k$ recommendations over $\traininter$. In other words, for each observed value of $\traininter$, we measure which fraction of items in the top-$k$ recommendations have fewer than $\traininter$ training interactions. Furthermore, we compute the catalog coverage~\cite{gunawardana2022evalrs_handbook}, \ie the fraction of distinct items that appear at least once in the top-$k$ recommendation lists.
    
    \subsection{Sequential Backbones. }We employ two state-of-the-art sequential, transformer-based RS as backbones to \ourapproach.
    \begin{description}
        \item[\bert{}~\cite{sun2019bert4rec}]: This model is based on the BERT architecture~\cite{devlin2019bert} and consists of a transformer  optimized through bidirectional self-attention.
        \item[\sasrec{}~\cite{kang2018sasrec}]: This model's architecture consists of a transformer optimized through causal self-attention.
    \end{description}
    We select these backbone sequential RS since they are often used in industrial scenarios and, more importantly, demonstrate state-of-the-art performance across several recommendation domains and evaluation setups~\cite{klenitskiy2023sasrec,koneru2024sasrec_zdf,betello2026sequential_reproducibility}.
    \subsection{Non-Sequential Baselines. }
    We compare overall model performance with well-established collaborative filtering RS, covering three different RS paradigms.
    \begin{description}
        \item[\multvae{}~\cite{liang2018multvae}] is based on a variational autoencoder architecture applied to the user interaction profile, and is considered among the strongest non-sequential neural-network-based architectures. 
        \item[\iknn{}~\cite{deshpande2004itemknn}] provides recommendations based on a item-nearest-neighbor approach, computing similarity based on the item interaction profile; it has been shown to often outperform neural-network-based approaches  despite its simplicity~\cite{dacrema2019are_we_really_making_progress}.
        \item[\als{}~\cite{hu2008als}] is a  matrix-factorization-based RS  that updates the user and item matrices by alternately holding one matrix fixed and updating the other, and is included as factorization-based baseline.
    \end{description}
    
\section{Results}
\newcommand{\res}[2]{$#1_{\scriptscriptstyle #2}$}
\newcommand{\bres}[2]{$\mathbf{#1}_{\scriptscriptstyle #2}$}

\begin{table*}[t]
\centering
\caption{Test HR@10 before and after swapping the last input item at
inference time, overall and for sequences ending with cold items
($\traininter\leq10$). CF baselines (bottom block of each dataset) do
not support swap inference or cold-item evaluation. Bold: best per
dataset and column; subscript: standard error.}
\label{tab:consolidated}
\begin{tabular}{ll rr rr}
\toprule
& & \multicolumn{2}{c}{Overall}
  & \multicolumn{2}{c}{Cold ($\traininter\leq10$)} \\
\cmidrule(lr){3-4}\cmidrule(lr){5-6}
\textbf{Dataset} & \textbf{Model}
  & no swap & swap & no swap & swap \\
\midrule
\multirow{7}{*}{\onion}
 & \sasrec   & \bres{.3183}{.0039} & \res{.0422}{.0017} & \res{.2269}{.0386} & \res{.1176}{.0297} \\
 & \swaprec  & \res{.3104}{.0039} & \bres{.2404}{.0036} & \bres{.2353}{.0390} & \bres{.1849}{.0357} \\
 & \bert     & \res{.2838}{.0038} & \res{.1402}{.0029} & \res{.2269}{.0386} & \res{.1597}{.0337} \\
 & \swapbert & \res{.2830}{.0038} & \res{.2009}{.0034} & \res{.2185}{.0380} & \res{.1597}{.0337} \\
\cmidrule(l){2-6}
 & \multvae  & \res{.1092}{.0026} & --- & --- & --- \\
 & \iknn     & \res{.0577}{.0019} & --- & --- & --- \\
 & \als      & \res{.0510}{.0019} & --- & --- & --- \\
\midrule
\multirow{7}{*}{\amazon}
 & \sasrec   & \res{.0517}{.0029} & \res{.0098}{.0013} & \res{.0095}{.0018} & \res{.0026}{.0009} \\
 & \swaprec  & \bres{.0518}{.0029} & \bres{.0454}{.0027} & \bres{.0099}{.0018} & \bres{.0082}{.0016} \\
 & \bert     & \res{.0510}{.0029} & \res{.0262}{.0021} & \res{.0086}{.0017} & \res{.0046}{.0012} \\
 & \swapbert & \res{.0511}{.0029} & \res{.0322}{.0023} & \res{.0089}{.0017} & \res{.0049}{.0013} \\
\cmidrule(l){2-6}
 & \multvae  & \res{.0010}{.0004} & --- & --- & --- \\
 & \iknn     & \res{.0071}{.0011} & --- & --- & --- \\
 & \als      & \res{.0215}{.0019} & --- & --- & --- \\
\midrule
\multirow{7}{*}{\movielens}
 & \sasrec   & \res{.1185}{.0020} & \res{.0524}{.0014} & \bres{.0098}{.0069} & \res{.0196}{.0097} \\
 & \swaprec  & \bres{.1202}{.0020} & \bres{.1102}{.0020} & \res{.0049}{.0049} & \res{.0147}{.0084} \\
 & \bert     & \res{.0978}{.0018} & \res{.0799}{.0017} & \res{.0049}{.0049} & \bres{.0245}{.0109} \\
 & \swapbert & \res{.0996}{.0019} & \res{.0967}{.0018} & \bres{.0098}{.0069} & \bres{.0245}{.0109} \\
\cmidrule(l){2-6}
 & \multvae  & \res{.0408}{.0013} & --- & --- & --- \\
 & \iknn     & \res{.0504}{.0014} & --- & --- & --- \\
 & \als      & \res{.0363}{.0012} & --- & --- & --- \\
\bottomrule
\end{tabular}
\end{table*}

    \paragraph{RQ1: Overall performance.} We start by evaluating the overall model performance on the full dataset without inference-time swaps. The results are reported on the left side of Table~\ref{tab:consolidated}
    . 
    
    All sequential models (\sasrec, \bert, with and without \ourapproach) reach higher HR than non-sequential models (\multvae, \iknn, \als) on all datasets. For \movielens and \amazon, this is in agreement with previous observations~\cite{klenitskiy2024does_it_look_sequential} reporting that for these datasets, interaction timestamps are informative for recommendation. For \onion, we 
    attribute this to the importance of sequentiality in music streaming~\cite{abbattista2024sequential_music,moscati2023actr_sequential_music,seshadri2024sequential_music}. We 
    observe that \sasrec variants outperform both \bert variants on all dataset, also in agreement with previous work~\cite{klenitskiy2023sasrec}. 

    Finally, we observe that \swaprec reaches the best performance on two datasets, indicating that our \swaprec, designed for item cold-start, improves recommendation accuracy even in a standard, warm scenario.
    
    \paragraph{RQ2: Robustness to Cold-Item Clicks.} 
    \begin{table*}[t]
\centering
\caption{HR@10 for different ways of handling strict-cold items:
swap (replace with $\phi(s_k)$), init (randomly initialize ID
embedding), and drop (remove from sequence). Evaluation restricted to
input sequences of at least two items. Bold: best per dataset;
subscript: standard error.}
\label{tab:alo_combined}
\begin{tabular}{ll Hrrr}
\toprule
\textbf{Dataset} & \textbf{Model} & no swap & swap & init & drop \\
\midrule
\multirow{4}{*}{\onion}
 & \bert     & \res{.2841}{.0038} & \res{.1407}{.0029} & \res{.1646}{.0031} & \res{.2112}{.0034} \\
 & \swapbert & \res{.2832}{.0038} & \res{.2015}{.0034} & \res{.1735}{.0032} & \res{.2124}{.0035} \\
 & \sasrec   & \res{.3186}{.0039} & \res{.0423}{.0017} & \res{.0681}{.0021} & \res{.2197}{.0035} \\
 & \swaprec  & \res{.3107}{.0039} & \bres{.2405}{.0036} & \res{.0813}{.0023} & \res{.2097}{.0034} \\
\midrule
\multirow{4}{*}{\amazon}
 & \bert     & \res{.0552}{.0066} & \res{.0360}{.0054} & \res{.0310}{.0050} & \res{.0301}{.0049} \\
 & \swapbert & \res{.0552}{.0066} & \res{.0335}{.0052} & \res{.0293}{.0049} & \res{.0310}{.0050} \\
 & \sasrec   & \res{.0569}{.0067} & \res{.0126}{.0032} & \res{.0033}{.0017} & \res{.0301}{.0049} \\
 & \swaprec  & \res{.0577}{.0068} & \bres{.0494}{.0063} & \res{.0126}{.0032} & \res{.0310}{.0050} \\
\midrule
\multirow{4}{*}{\movielens}
 & \bert     & \res{.0978}{.0019} & \res{.0799}{.0017} & \res{.0804}{.0017} & \res{.0714}{.0016} \\
 & \swapbert & \res{.0996}{.0019} & \res{.0967}{.0019} & \res{.0802}{.0017} & \res{.0740}{.0017} \\
 & \sasrec   & \res{.1185}{.0020} & \res{.0524}{.0014} & \res{.0701}{.0016} & \res{.0871}{.0018} \\
 & \swaprec  & \res{.1202}{.0021} & \bres{.1102}{.0020} & \res{.0747}{.0017} & \res{.0894}{.0018} \\
\bottomrule
\end{tabular}
\end{table*}  
    \begin{figure}[t]
\centering

\pgfplotsset{
    coldstart/.style={
        ybar,
        bar width=4.5pt,
        width=\linewidth,
        height=4.4cm,
        enlarge x limits=0.12,
        ymin=0,
        ylabel style={font=\small},
        xtick=data,
        xticklabels={,,,},               
        xticklabel style={font=\small},
        yticklabel style={font=\small},
        xlabel={},
        legend style={
            at={(0.5,1.06)},
            anchor=south,
            legend columns=2,
            font=\small,
            column sep=8pt,
            draw=none,
        },
        legend cell align=left,
        ymajorgrids=true,
        grid style={thin, gray!25},
        axis line style={gray!60},
        tick style={draw=none},
        error bars/y dir=both,
        error bars/y explicit,
        error bars/error bar style={line width=0.5pt, gray!70},
        error bars/error mark options={line width=0.5pt, gray!70, mark size=1.5pt},
    }
}

\begin{subfigure}[t]{\linewidth}
\centering
\begin{tikzpicture}
\begin{axis}[
    coldstart,
    ymax=0.40,
    ytick={0.00,0.10,0.20,0.30,0.40},
    ylabel={HR@10},
]
\addplot[fill=blue!70!black, draw=none] plot[error bars/.cd, y explicit] coordinates {
    (1,0.2269)+-(0,0.0386) (2,0.2057)+-(0,0.0342) (3,0.2717)+-(0,0.0109) (4,0.3269)+-(0,0.0043)};
\addlegendentry{No swap}
\addplot[fill=blue!30, draw=none] plot[error bars/.cd, y explicit] coordinates {
    (1,0.1176)+-(0,0.0297) (2,0.0284)+-(0,0.0140) (3,0.0360)+-(0,0.0046) (4,0.0424)+-(0,0.0018)};
\addlegendentry{Swap}
\addplot[fill=orange!80!black, draw=none] plot[error bars/.cd, y explicit] coordinates {
    (1,0.2353)+-(0,0.0390) (2,0.2199)+-(0,0.0350) (3,0.2675)+-(0,0.0108) (4,0.3181)+-(0,0.0042)};
\addlegendentry{No swap w. \ourapproach}
\addplot[fill=orange!40, draw=none] plot[error bars/.cd, y explicit] coordinates {
    (1,0.1849)+-(0,0.0357) (2,0.1206)+-(0,0.0275) (3,0.1542)+-(0,0.0088) (4,0.2542)+-(0,0.0040)};
\addlegendentry{Swap w. \ourapproach}
\draw[dashed, thick, blue!70!black] (axis cs:0.55,0.3183)--(axis cs:4.45,0.3183);
\draw[dotted, thick, orange!80!black] (axis cs:0.55,0.3104)--(axis cs:4.45,0.3104);
\node[font=\scriptsize,text=blue!70!black,anchor=south east] at (axis cs:1.55,0.3183) {\sasrec\ overall};
\node[font=\scriptsize,text=orange!80!black,anchor=north east] at (axis cs:1.55,0.3104) {\swaprec\ overall};
\end{axis}
\end{tikzpicture}
\caption{\onion}
\end{subfigure}

\vspace{4pt}
\begin{subfigure}[t]{\linewidth}
\centering
\begin{tikzpicture}
\begin{axis}[
    coldstart,
    ymax=0.22,
    ytick={0.00,0.10,0.20},
    ylabel={HR@10},
]
\addplot[fill=blue!70!black, draw=none] plot[error bars/.cd, y explicit] coordinates {
    (1,0.0095)+-(0,0.0018) (2,0.0447)+-(0,0.0071) (3,0.0933)+-(0,0.0080) (4,0.1643)+-(0,0.0140)};
\addplot[fill=blue!30, draw=none] plot[error bars/.cd, y explicit] coordinates {
    (1,0.0026)+-(0,0.0009) (2,0.0059)+-(0,0.0026) (3,0.0191)+-(0,0.0038) (4,0.0283)+-(0,0.0062)};
\addplot[fill=orange!80!black, draw=none] plot[error bars/.cd, y explicit] coordinates {
    (1,0.0099)+-(0,0.0018) (2,0.0435)+-(0,0.0070) (3,0.0933)+-(0,0.0080) (4,0.1657)+-(0,0.0140)};
\addplot[fill=orange!40, draw=none] plot[error bars/.cd, y explicit] coordinates {
    (1,0.0082)+-(0,0.0016) (2,0.0329)+-(0,0.0061) (3,0.0772)+-(0,0.0074) (4,0.1615)+-(0,0.0139)};
\draw[dashed, thick, blue!70!black] (axis cs:0.55,0.0517)--(axis cs:4.45,0.0517);
\draw[dotted, thick, orange!80!black] (axis cs:0.55,0.0518)--(axis cs:4.45,0.0518);
\node[font=\scriptsize,text=blue!70!black,anchor=south west] at (axis cs:0.95,0.0517) {\sasrec\ overall};
\node[font=\scriptsize,text=orange!80!black,anchor=north west] at (axis cs:0.95,0.0518) {\swaprec\ overall};
\end{axis}
\end{tikzpicture}
\caption{\amazon}
\end{subfigure}

\vspace{4pt}
\begin{subfigure}[t]{\linewidth}
\centering
\begin{tikzpicture}
\begin{axis}[
    coldstart,
    ymax=0.22,
    ytick={0.00,0.10,0.20},
    ylabel={HR@10},
    xticklabels={$\leq$10, 11--20, 21--100, $\geq$101},
    xlabel={Interactions in train set},
    xlabel style={font=\small, yshift=-2pt},
]
\addplot[fill=blue!70!black, draw=none] plot[error bars/.cd, y explicit] coordinates {
    (1,0.0098)+-(0,0.0069) (2,0.0237)+-(0,0.0117) (3,0.0427)+-(0,0.0084) (4,0.1220)+-(0,0.0021)};
\addplot[fill=blue!30, draw=none] plot[error bars/.cd, y explicit] coordinates {
    (1,0.0196)+-(0,0.0097) (2,0.0237)+-(0,0.0117) (3,0.0222)+-(0,0.0061) (4,0.0536)+-(0,0.0015)};
\addplot[fill=orange!80!black, draw=none] plot[error bars/.cd, y explicit] coordinates {
    (1,0.0049)+-(0,0.0049) (2,0.0178)+-(0,0.0102) (3,0.0392)+-(0,0.0080) (4,0.1239)+-(0,0.0021)};
\addplot[fill=orange!40, draw=none] plot[error bars/.cd, y explicit] coordinates {
    (1,0.0147)+-(0,0.0084) (2,0.0296)+-(0,0.0131) (3,0.0341)+-(0,0.0075) (4,0.1134)+-(0,0.0021)};
\draw[dashed, thick, blue!70!black] (axis cs:0.55,0.1185)--(axis cs:4.45,0.1185);
\draw[dotted, thick, orange!80!black] (axis cs:0.55,0.1202)--(axis cs:4.45,0.1202);
\node[font=\scriptsize,text=blue!70!black,anchor=south west] at (axis cs:0.95,0.1185) {\sasrec\ overall};
\node[font=\scriptsize,text=orange!80!black,anchor=north west] at (axis cs:0.95,0.1202) {\swaprec\ overall};
\end{axis}
\end{tikzpicture}
\caption{\movielens}
\end{subfigure}

\caption{HR@10 by number of interactions of the last input item. Error bars: $\pm 1$ SE. Dashed lines: overall HR@10.}
\label{fig:coldstart_bins}
\end{figure}
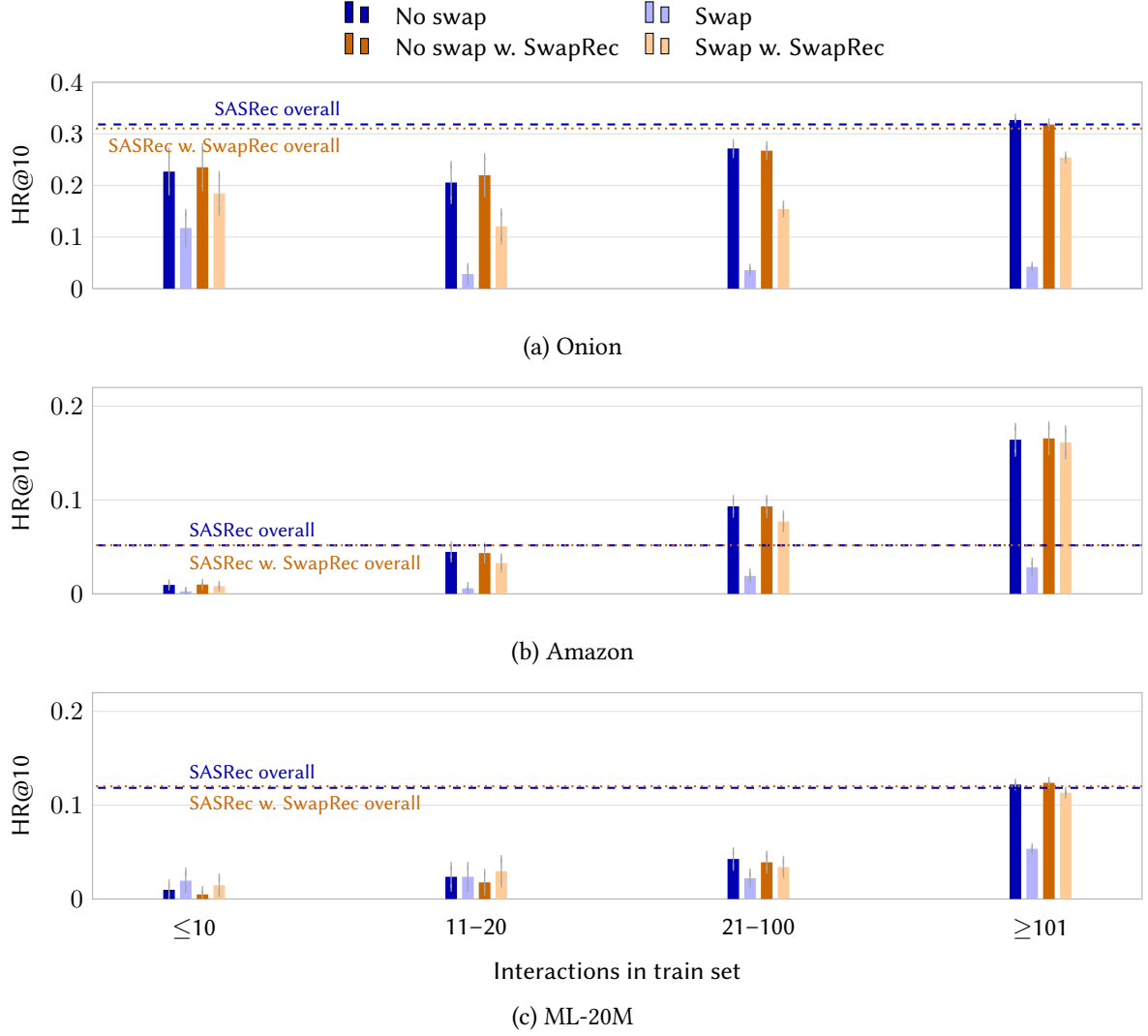
    Moving on to the main focus of this work, \ie item cold-start recommendation, the right side of Table~\ref{tab:consolidated} shows results for sequential models and includes the performance when the last item in the input sequence is swapped with its most-similar one. The columns ``\textit{Cold}'' refer to test sequences ending with a cold item with $\traininter < \ncold$. We first observe that all models experience a lower performance when the last item is cold (``\textit{Overall-no swap}'' vs. ``\textit{Cold-no swap}''), demonstrating the expected performance deterioration in cold-start scenarios. We see that in both cases (``\textit{Overall}'' and ``\textit{Cold}''), when looking at  model performance after applying the swap at inference time (``\textit{swap}''), models experience a performance deterioration. This is expected, especially on the full dataset, where the learned token embedding of the last item is more representative of the session than the swapped one. However, the performance deterioration is less remarked when models were trained with \ourapproach, demonstrating that \ourapproach makes models more robust to the item swap heuristic often used to address cold-start scenarios. 
    Table~\ref{tab:alo_combined} reports results in the strict item cold-start scenarios, \ie when $\traininter = 0$. For each sequential model, we include three ways of handling strict-cold-start items: keeping the randomly initialized token embedding (``\textit{init}''), dropping the strict-cold-start item from the input (``\textit{drop}''), and swapping it at inference time with its most similar warm item (``\textit{swap}'').

    We observe that for models without \ourapproach, swapping the cold item with the warm one at inference time can result in a worse performance than other ways of addressing item cold-start. This is the case especially on \onion, and we attribute this to the fact that since music streaming often occurs in sessions, sequences tend to be longer as compared to other domains such as movie streaming or online shopping, and hence the model can ``fall back'' on the information from the other items in the input sequence. 
    
    We also observe that models trained with \ourapproach often reach a better performance than the model trained without, indicating the effectiveness of \ourapproach to render models more robust to item cold-start \textit{irrespective} of the strategy used to address it; we attribute this to the ID embedding updates of cold items that happen due to swaps during training. As expected, \ourapproach also allows models to reach their best item-cold-start performance when swapping is used at inference time. 
    
    To summarize, the results in Tables~\ref{tab:consolidated} and~\ref{tab:alo_combined} indicate that inference-time swaps of cold items with their most content-similar ones can negatively impact model performance, and that by leveraging these swaps during training, \ourapproach renders models more robust to item cold-start. 
    
    We further analyze the impact of $\traininter$. For conciseness, we carry out the analysis on the best performing sequential model (\sasrec), with and without \ourapproach, as summarized in Figure~\ref{fig:coldstart_bins}. As expected, we see that all models reach a lower performance for sequences ending with cold items ($\traininter < 10$), demonstrating that such sequences can harm real-time recommendation. %
    In almost all cases, models experience a performance deterioration when swapping the last item in the sequence, and we attribute this to the fact that, considering the short-head long-tail distribution of interaction data, items are likely to be mapped to cold items when the swaps are applied at inference time. 
    
    When applying \ourapproach, the deterioration mostly decreases as $\traininter$ decreases; this is a further indication of \ourapproach's effectiveness in handling cold-start items: \ourapproach allows having a valid representation for cold items, avoiding staticity when interactions with these items occur. 
    
    Overall, we see that when trained with \ourapproach, the performance deterioration in presence of interactions with cold items is less pronounced, irrespective of the number of training interactions of the swapped item and of the strategy used to handle those interactions. 
    \paragraph{RQ3: Cold Items Recommended.} RQ3 focuses on \ourapproach's positive effect for content producers, \ie its ability to expose cold items more often in recommendation lists.
    
    \begin{table}[t]
\centering
\caption{Cold-start impact metrics for \amazon. Bold coverage values highlight \ourapproach relative to \sasrec.}
\label{tab:coldstart-impact}
\begin{tabular}{lHHHrrHHH}
\toprule
Model & Strategy & Swap variant & Warm-only $n$ & Cold-\% [95\% CI] & Coverage & $n_{\text{distinct}}$/catalog & Cold-target HR@10 [95\% CI] & $n$ (cold-target) \\
\midrule
\sasrec  & swap     & vocab\_restricted & 4799 & 30.7\% [30.3, 31.1] & 0.5779          & 4106/7105 & 0.0000 [0.0, 0.0] & 78 \\
\ourapproach & swap     & vocab\_restricted & 4799 & 31.8\% [31.4, 32.2] & {0.6322} & 4492/7105 & 0.0000 [0.0, 0.0] & 78 \\
\bottomrule
\end{tabular}%
\end{table}

\begin{figure}[t]
  \centering
  \includegraphics[width=\linewidth]{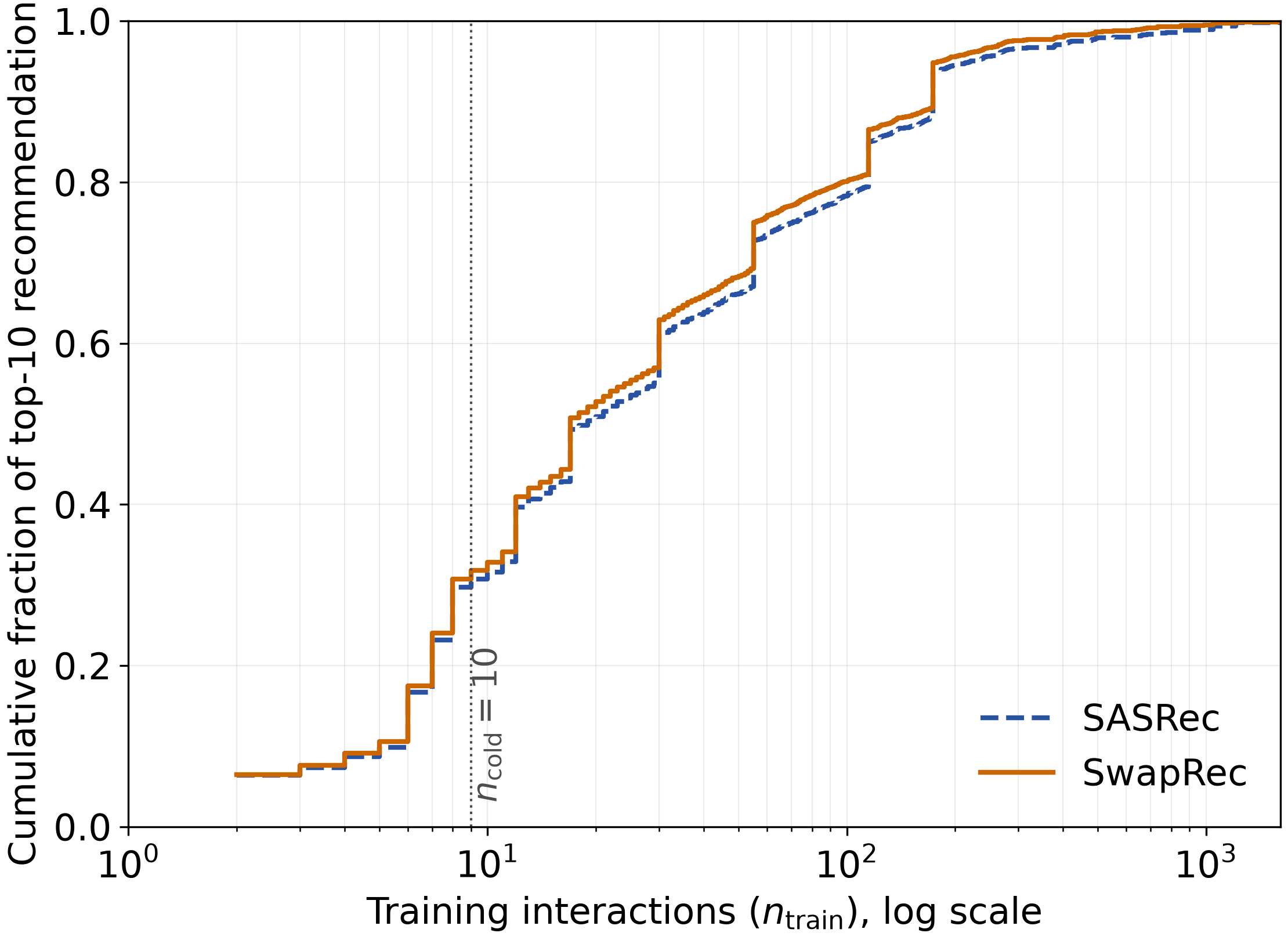}
  \caption{Empirical CDF of training-set frequency ($n_{\mathrm{train}}$)
  among top-10 recommended items for \amazon test sequences. \ourapproach's curve is always above \sasrec, indicating that \ourapproach recommends less popular items.}
  \label{fig:amazon_ntrain_cdf}
\end{figure}

    Figure~\ref{fig:amazon_ntrain_cdf} shows the empirical CDF of $\traininter$ appearing in the aggregated top-10 recommendation lists for \amazon test sequences. For the same aggregated top-10 recommendations, Table~\ref{tab:coldstart-impact} shows the percentage of cold items with $\traininter<\ncold$, as well as the catalog coverage, for both the strongest comparison model (\sasrec) and its variant leveraging \ourapproach. \ourapproach's curve is always above \sasrec, indicating that \ourapproach recommends less popular items.. We see that \ourapproach improves on both the percentage of cold items recommended, as well as the overall catalog coverage. Figure~\ref{fig:coldstart_bins} shows the CDF of top-$k$ \amazon recommendations over $\traininter$. We observe that \ourapproach's CDF is always above its underlying backbone \sasrec, indicating that \ourapproach leads to less popular recommendations, and that the higher percentage of cold recommendations is not an artifact of how the threshold $\ncold$ is defined. 
    
    We attribute both the larger catalog coverage and the higher percentage of cold-items recommendations to the fact that through training-time swaps, \ourapproach leads to more frequent updates of cold item ID embeddings, as compared to the underlying backbone. 

\section{Conclusions and Future Work}
    This work focused on mitigating the negative impact of clicks on  cold items on sequential recommendation. We showed that exchanging cold items with their most similar item, an approach used to ensure real-time personalization in industry environments, is only effective if also applied during training. We hence proposed \ourapproach, which applies training-time swapping to transformer-based sequential recommenders, demonstrating that \ourapproach mitigates the performance deterioration in item cold-start and leads to a larger percentage of cold-item occurrences in recommendation lists. 
    
    There are several limitations to our work. 
    We limited item swaps during evaluation to the last item in the sequence. We developed  \ourapproach limiting the map exclusively to the most similar  warm item. We did not investigate the impact of 
    multiple inference-time swaps, different swap positions, or different swap maps. We think that the use of contrastive losses -- like in natural language processing for synonyms -- could further improve models' robustness to swaps.  Finally, we did not investigate the impact of \ourapproach on the structure of the ID embedding space. 
    
    We leave these extensions for future research.

\section*{Declaration on Generative AI}
    The authors used Claude Code during the development of \ourapproach and to generate the figures of the paper. After using these tools, the authors reviewed and edited the code as needed and take full responsibility for the content produced.

\bibliography{sample-ceur}

@INPROCEEDINGS{hu2008als,
  author={Hu, Yifan and Koren, Yehuda and Volinsky, Chris},
  booktitle={Proc. of IEEE ICDM}, 
  title={Collaborative Filtering for Implicit Feedback Datasets}, 
  year={2008},
  }

@article{
yang2025liger,
title={Unifying Generative and Dense Retrieval for Sequential Recommendation},
author={Liu Yang and Fabian Paischer and Kaveh Hassani and Jiacheng Li and Shuai Shao and Zhang Gabriel Li and Yun He and Xue Feng and Nima Noorshams and Sem Park and Bo Long and Robert D Nowak and Xiaoli Gao and Hamid Eghbalzadeh},
journal={Transactions on Machine Learning Research},
year={2025},
}

@inproceedings{lfm2b,
	title        = {LFM-2b: A Dataset of Enriched Music Listening Events for Recommender Systems Research and Fairness Analysis},
	author       = {Schedl, Markus and Brandl, Stefan and Lesota, Oleg and Parada-Cabaleiro, Emilia and Penz, David and Rekabsaz, Navid},
	booktitle    = {Proc. of ACM CHIIR},
	location     = {Regensburg, Germany},
	year         = 2022,
}

@inproceedings{pons2018atscale,
  title={End-to-end learning for music audio tagging at scale},
  author={Pons, Jordi and Nieto, Oriol and Prockup, Matthew and Schmidt, Erik M. and Ehmann, Andreas F. and Serra, Xavier},
  booktitle={Proc. of ISMIR},
  year={2018},
}

@inproceedings{pons2019musicnn,
  title={musicnn: pre-trained convolutional neural networks for music audio tagging},
  author={Pons, Jordi and Serra, Xavier},
  booktitle={Late-breaking/demo session in 20th International Society for Music Information Retrieval Conference (LBD-ISMIR2019)},
  year={2019},
}

@article{peintner2025srgnnemo,
    title = {Nuanced Music Emotion Recognition via a Semi-Supervised Multi-Relational Graph Neural Network},
    author = {Peintner, Andreas and {Moscati}, Marta and Kinoshita, Yu and Vogl, Richard and Knees, Peter and Schedl, Markus and Strauss, Hannah and Zenter, Marcel and Zangerle, Eva},
    journal = {Trans. of ISMIR},
    publisher = {Association for Computing Machinery},
    volume = {8},
    year = {2025}
}

@inproceedings{abbattista2024sequential_music,
author = {Abbattista, Davide and Anelli, Vito Walter and Di Noia, Tommaso and Macdonald, Craig and Petrov, Aleksandr Vladimirovich},
title = {Enhancing Sequential Music Recommendation with Personalized Popularity Awareness},
year = {2024},
booktitle = {Proc. of ACM RecSys},
location = {Bari, Italy},
}

@misc{spillo2026movielens,
	title = {Binge {Watch}: {Reproducible} {Multimodal} {Benchmarks} {Datasets} for {Large}-{Scale} {Movie} {Recommendation} on {MovieLens}-{10M} and {20M}},
	author = {Spillo, Giuseppe and Petruzzelli, Alessandro and Musto, Cataldo and Gemmis, Marco de and Lops, Pasquale and Semeraro, Giovanni},
	year = {2026},
      eprint={2602.15505},
      archivePrefix={arXiv},
      primaryClass={cs.LG},
}

@inproceedings{santana2020music4all,
  title={Music4all: A new music database and its applications},
  author={Santana, Igor Andr{\'e} Pegoraro and Pinhelli, Fabio and Donini, Juliano and Catharin, Leonardo and Mangolin, Rafael Biazus and Feltrim, Val{\'e}ria Delisandra and Domingues, Marcos Aur{\'e}lio and others},
  booktitle={Proc. of IWSSIP},
  year={2020},
}

@inproceedings{gusak2025time_to_split,
author = {Gusak, Danil and Volodkevich, Anna and Klenitskiy, Anton and Vasilev, Alexey and Frolov, Evgeny},
title = {Time to Split: Exploring Data Splitting Strategies for Offline Evaluation of Sequential Recommenders},
year = {2025},
booktitle = {Proc. of ACM RecSys},
location = {Prague, Czech Republic},
}

@inproceedings{spillo2023combining,
author = {Spillo, Giuseppe and Musto, Cataldo and Polignano, Marco and Lops, Pasquale and de Gemmis, Marco and Semeraro, Giovanni},
title = {Combining Graph Neural Networks and Sentence Encoders for Knowledge-aware Recommendations},
year = {2023},
booktitle = {Proc. of ACM UMAP},
location = {Limassol, Cyprus},
series = {UMAP '23}
}

@inproceedings{ganhoer_moscati2024sibrar,
    title = {A Multimodal Single-Branch Embedding Network for Recommendation in Cold-Start and Missing Modality Scenarios},
    author = {Ganhör*, Christian and Moscati*, Marta and Hausberger, Anna and Nawaz, Shah and Schedl, Markus},
    booktitle = {Proc. of ACM RecSys},
    location = {Bari, Italy},
    year = {2024}
}

@article{Melchiorre2021RSFairness,
    title = {Investigating Gender Fairness of Recommendation Algorithms in the Music Domain},
    author = {Melchiorre, Alessandro B. and Rekab-saz, Navid and Parada-Cabaleiro, Emilia and Brandl, Stefan and Lesota, Oleg and Schedl, Markus},
    journal = {Information Processing and Management},
    volume = {58},
    number = {5},
    year = {2021}
}

@article{Ganhoer2025TORS_SiBraR,
    title = {Single-Branch Network Architectures to Close the Modality Gap in Multimodal Recommendation},
    author = {Ganhör, Christian and Moscati, Marta and Hausberger, Anna and Nawaz, Shah and Schedl, Markus},
    journal = {ACM Transactions on Recommender Systems},
    year = {2025}
}

@inproceedings{li2026diffusion,
      title={Adaptive Autoguidance for Item-Side Fairness in Diffusion Recommender Systems}, 
      author={Zihan Li and Gustavo Escobedo and Marta Moscati and Oleg Lesota and Markus Schedl},
      year={2026},
	booktitle    = {Proc. of ACM SIGIR},
	location     = {Melbourne, Australia},
	year         = 2026,
}

@article{malitesta2024formal,
author = {Malitesta, Daniele and Cornacchia, Giandomenico and Pomo, Claudio and Merra, Felice Antonio and Di Noia, Tommaso and Di Sciascio, Eugenio},
title = {Formalizing Multimedia Recommendation through Multimodal Deep Learning},
year = {2024},
journal = {Transactions on Recommender Systems},
}

@book{ricci2022rshandbook,
    editor="Ricci, Francesco
    and Rokach, Lior
    and Shapira, Bracha",
    title="Recommender Systems Handbook",
    year="2022",
    publisher="Springer US",
    address="New York, NY",
}

@inproceedings{lichtenberg2025denserec,
  author       = {Lichtenberg, Jan Malte and De Candia, Antonio and Ruffini, Matteo},
  title        = {DenseRec: Revisiting Dense Content Embeddings for Sequential Transformer-based Recommendation},
  booktitle    = {Proc. of EARL Workshop at {ACM} RecSys},
  year         = {2025},
location = {Prague, Czech Republic},
}

@incollection{gunawardana2022evalrs_handbook,
    author="Asela Gunawardana and Guy Shani and Sivan Yogev",
    editor="Ricci, Francesco
    and Rokach, Lior
    and Shapira, Bracha",
    title="Evaluating Recommender Systems",
    bookTitle="Recommender Systems Handbook",
    year="2022",
    publisher="Springer US",
    address="New York, NY",
    pages="547--602",
}

@inproceedings{oh2024recency,
title={Measuring Recency Bias In Sequential Recommendation Systems}, 
      author={Jeonglyul Oh and Sungzoon Cho},
  booktitle    = {Proc. of CONSEQUENCES Workshop at {ACM} RecSys},
  year         = {2024},
location = {Bari, Italy},
}

@inproceedings{gusak2025recsys_dish_served_warm,
author = {Gusak, Danil and Sukhorukov, Nikita and Frolov, Evgeny},
title = {Recommendation Is a Dish Better Served Warm},
year = {2025},
booktitle = {Proc. of ACM RecSys},
location = {Prague, Czech Republic},
}

@misc{collins2026ensrec,
      title={Exploiting ID-Text Complementarity via Ensembling for Sequential Recommendation}, 
      author={Liam Collins and Bhuvesh Kumar and Clark Mingxuan Ju and Tong Zhao and Donald Loveland and Leonardo Neves and Neil Shah},
      year={2026},
      eprint={2512.17820},
      archivePrefix={arXiv},
      primaryClass={cs.LG},
}

@article{harper2015movielens20m,
author = {Harper, F. Maxwell and Konstan, Joseph A.},
title = {The MovieLens Datasets: History and Context},
year = {2015},
volume = {5},
number = {4},
journal = {ACM Transactions on Interactive Intelligent Systems},
articleno = {19},
}

@inproceedings{lichtenberg2025daquamrec,
  author       = {Lichtenberg, Jan Malte and Ruffini, Matteo},
  title        = {Sequential Recommenders and Multimodal Inputs: Mitigating Data Quality Issues in Industry-Scale Recommenders},
  booktitle    = {Proc. of DaQuaMRec Workshop at {ACM} RecSys},
  year         = {2025},
location = {Prague, Czech Republic},
}

@inproceedings{reimers2019sentence_bert,
  title = "Sentence-BERT: Sentence Embeddings using Siamese BERT-Networks",
  author = "Reimers, Nils and Gurevych, Iryna",
  booktitle = "Proc. of EMNLP",
  year = 2019,
}

@inproceedings{dacrema2019are_we_really_making_progress,
author = {Ferrari Dacrema, Maurizio and Cremonesi, Paolo and Jannach, Dietmar},
title = {Are we really making much progress? A worrying analysis of recent neural recommendation approaches},
year = {2019},
booktitle = {Proc. of ACM RecSys Systems},
location = {Copenhagen, Denmark},
}

@inproceedings{klenitskiy2024does_it_look_sequential,
author = {Klenitskiy, Anton and Volodkevich, Anna and Pembek, Anton and Vasilev, Alexey},
title = {Does It Look Sequential? An Analysis of Datasets for Evaluation of Sequential Recommendations},
year = {2024},
booktitle = {Proc. of ACM RecSys},
numpages = {6},
location = {Bari, Italy},
series = {RecSys '24}
}

@article{deshpande2004itemknn,
author = {Deshpande, Mukund and Karypis, George},
title = {Item-based top-N recommendation algorithms},
year = {2004},
volume = {22},
number = {1},
journal = {ACM Transactions on Information Systems},
}

@inproceedings{liang2018multvae,
author = {Liang, Dawen and Krishnan, Rahul G. and Hoffman, Matthew D. and Jebara, Tony},
title = {Variational Autoencoders for Collaborative Filtering},
year = {2018},
booktitle = {Proc. of WWW},
location = {Lyon, France},
}

@inproceedings{grbovic2018real_time_personalization_airbnb,
author = {Grbovic, Mihajlo and Cheng, Haibin},
title = {Real-time Personalization using Embeddings for Search Ranking at Airbnb},
year = {2018},
booktitle = {Proc. of ACM SIGKDD},
location = {London, United Kingdom},
}

@inproceedings{seol2024proxy_alibaba,
author = {Seol, Jinseok and Gang, Minseok and Lee, Sang-goo and Park, Jaehui},
title = {Proxy-based Item Representation for Attribute and Context-aware Recommendation},
year = {2024},
booktitle = {Proc. of ACM WSDM},
location = {Merida, Mexico},
}

@inproceedings{singh2024semantic_ids,
author = {Singh, Anima and Vu, Trung and Mehta, Nikhil and Keshavan, Raghunandan and Sathiamoorthy, Maheswaran and Zheng, Yilin and Hong, Lichan and Heldt, Lukasz and Wei, Li and Tandon, Devansh and Chi, Ed and Yi, Xinyang},
title = {Better Generalization with Semantic IDs: A Case Study in Ranking for Recommendations},
year = {2024},
booktitle = {Proc. of ACM RecSys},
numpages = {6},
location = {Bari, Italy},
}

@inproceedings{rajput2023semantic_ids,
author = {Rajput, Shashank and Mehta, Nikhil and Singh, Anima and Keshavan, Raghunandan and Vu, Trung and Heidt, Lukasz and Hong, Lichan and Tay, Yi and Tran, Vinh Q. and Samost, Jonah and Kula, Maciej and Chi, Ed H. and Sathiamoorthy, Maheswaran},
title = {Recommender systems with generative retrieval},
year = {2023},
booktitle = {Proc. of NeurIPS},
location = {New Orleans, LA, USA},
}

@inproceedings{koneru2024sasrec_zdf,
author = {Koneru, Venkata Harshit and Neufeld, Xenija and Loth, Sebastian and Gr\"{u}n, Andreas},
title = {Enhancing Recommendation Quality of the SASRec Model by Mitigating Popularity Bias},
year = {2024},
booktitle = {Proc. of ACM RecSys},
location = {Bari, Italy},
series = {RecSys '24}
}

@inproceedings{moscati2023actr_sequential_music,
author = {Moscati, Marta and Wallmann, Christian and Reiter-Haas, Markus and Kowald, Dominik and Lex, Elisabeth and Schedl, Markus},
title = {Integrating the ACT-R Framework with Collaborative Filtering for Explainable Sequential Music Recommendation},
year = {2023},
booktitle = {Proc. of ACM RecSys},
location = {Singapore, Singapore},
}

@inproceedings{seshadri2024sequential_music,
author = {Seshadri, Pavan and Shashaani, Shahrzad and Knees, Peter},
title = {Enhancing Sequential Music Recommendation with Negative Feedback-informed Contrastive Learning},
year = {2024},
booktitle = {Proc. of ACM RecSys},
location = {Bari, Italy},
}

@inproceedings{koneru2025sasrec_zdf,
author = {Koneru, Venkata Harshit and Neufeld, Xenija and Loth, Sebastian and Gr\"{u}n, Andreas},
title = {SASRec in Action: Real-World Adaptations for ZDF Streaming Service},
year = {2025},
booktitle = {Proc. of ACM RecSys},
location = {Prague, Czech Republic},
series = {RecSys '25}
}

@inproceedings{klenitskiy2023sasrec,
author = {Klenitskiy, Anton and Vasilev, Alexey},
title = {Turning Dross Into Gold Loss: is BERT4Rec really better than SASRec?},
year = {2023},
booktitle = {Proc. of ACM RecSys},
location = {Singapore, Singapore},
series = {RecSys '23}
}

@inproceedings{pembec2025letitgo,
author = {Pembek, Anton and Fatkulin, Artem and Klenitskiy, Anton and Vasilev, Alexey},
title = {Let It Go? Not Quite: Addressing Item Cold Start in Sequential Recommendations with Content-Based Initialization},
year = {2025},
booktitle = {Proc. of ACM RecSys},
location = {Prague, Czech Republic},
series = {RecSys '25}
}

@inproceedings{wei2021clcrec,
author = {Wei, Yinwei and Wang, Xiang and Li, Qi and Nie, Liqiang and Li, Yan and Li, Xuanping and Chua, Tat-Seng},
title = {Contrastive Learning for Cold-Start Recommendation},
year = {2021},
booktitle = {Proc. of ACM Multimedia},
pages = {5382--5390},
location = {Virtual Event, China},
}

@inproceedings{volkovs2017dropoutnet,
 author = {Volkovs, Maksims and Yu, Guangwei and Poutanen, Tomi},
 booktitle = {Proc. of NeurIPS},
 title = {DropoutNet: Addressing Cold Start in Recommender Systems},
 location = {Long Beach, CA, USA},
 year = {2017}
}

@inproceedings{kang2018sasrec,
  author={Kang, Wang-Cheng and McAuley, Julian},
  booktitle={Proc. of IEEE ICDM}, 
  title={Self-Attentive Sequential Recommendation}, 
  year={2018},
  }

@inproceedings{sun2019bert4rec,
author = {Sun, Fei and Liu, Jun and Wu, Jian and Pei, Changhua and Lin, Xiao and Ou, Wenwu and Jiang, Peng},
title = {BERT4Rec: Sequential Recommendation with Bidirectional Encoder Representations from Transformer},
year = {2019},
booktitle = {Proc. of ACM CIKM},
location = {Beijing, China},
}

@inproceedings{moscati2021onion,
  author    = {Marta Moscati and
               Emilia Parada{-}Cabaleiro and
               Yashar Deldjoo and
               Eva Zangerle and
               Markus Schedl},
  title     = {Music4All-Onion - {A} Large-Scale Multi-faceted Content-Centric Music
               Recommendation Dataset},
  booktitle = {Proc. of  {ACM} CIKM},
  location = {Atlanta, GA, USA},
  year      = {2022},
}

@inproceedings{ni2019amazonreviews,
    title = "Justifying Recommendations using Distantly-Labeled Reviews and Fine-Grained Aspects",
    author = "Ni, Jianmo  and
      Li, Jiacheng  and
      McAuley, Julian",
    booktitle = "Proc. of EMNLP",
    location = "Hong Kong, China",
    year = "2019",
}

@ARTICLE{betello2026sequential_reproducibility,
  author={Betello, Filippo and Purificato, Antonio and Siciliano, Federico and Trappolini, Giovanni and Bacciu, Andrea and Tonellotto, Nicola and Silvestri, Fabrizio},
  journal={IEEE Access}, 
  title={A Reproducible Analysis of Sequential Recommender Systems}, 
  year={2025},
  volume={13},
  }

@inproceedings{devlin2019bert,
  title={BERT: Pre-training of Deep Bidirectional Transformers for Language Understanding},
  author={Jacob Devlin and Ming-Wei Chang and Kenton Lee and Kristina Toutanova},
  booktitle={North American Chapter of the Association for Computational Linguistics},
  year={2019},
  url={https://api.semanticscholar.org/CorpusID:52967399}
}

@inproceedings{hou2024amazonreviews,
    title = "Bridging Language and Items for Retrieval and Recommendation: Benchmarking {LLM}s as Semantic Encoders",
    author = "Hou, Yupeng  and
      Li, Jiacheng  and
      Fu, Xiangjun  and
      He, Zhankui  and
      Yan, An  and
      Chen, Xiusi  and
      McAuley, Julian",
    booktitle = "Proc. of ACL",
    year = "2026",
    address = "San Diego, California, USA",
}

\end{document}